\documentclass[a4paper]{spie}  

\usepackage[]{graphicx}
\usepackage{amssymb,amsmath,psfrag,url}
\usepackage{url}
\urldef{\urldrlitho}{\url}{http://www.drlitho.com}
\urldef{\urljcmwave}{\url}{http://www.jcmwave.com}

\title{Benchmark of Rigorous Methods\\for Electromagnetic Field Simulations}

\author{
Sven Burger\supit{\,ab},
Lin Zschiedrich\supit{\,ab},
Frank Schmidt\supit{\,ab},\\
Peter Evanschitzky\supit{\,c},
Andreas Erdmann\supit{\,c}
\skiplinehalf
\supit{a}
Zuse Institute Berlin,
Takustra{\ss}e 7,
D\,--\,14\,195 Berlin,
Germany
\smallskip\\
\supit{b}
JCMwave GmbH,
Haarer Stra{\ss}e 14a,
D\,--\,85\,640 Putzbrunn,
Germany
\smallskip\\
\supit{c}
Fraunhofer Institute of Integrated Systems and Device Technology\\
(Fraunhofer IISB),
Schottkystra{\ss}e 10,
D\,--\,91\,058 Erlangen,
Germany
}

\authorinfo{
Corresponding author: S. Burger\\
URL: http://www.zib.de/Numerik/NanoOptics/\\
Email: burger@zib.de
}

\begin{document}
  \maketitle

\noindent
This paper will be published in Proc.~SPIE Vol. {\bf 7122}, 71221S 
(2008),  ({\it Photomask Technology 2008, Hiroichi Kawahira; Larry S. Zurbrick, Eds.}, 
Society of Photo-Optical Instrumentation Engineers, 2008)
and is made available 
as an electronic preprint with permission of SPIE. 
One print or electronic copy may be made for personal use only. 
Systematic or multiple reproduction, distribution to multiple 
locations via electronic or other means, duplication of any 
material in this paper for a fee or for commercial purposes, 
or modification of the content of the paper are prohibited.

\begin{abstract}

We have developed an interface which allows to perform rigorous electromagnetic field (EMF)
simulations with the simulator JCMsuite and subsequent aerial imaging and resist simulations
with the simulator Dr.LiTHO.
With the combined tools we investigate the convergence of near-field and far-field results
for different DUV masks.
We also benchmark results obtained with the waveguide-method EMF
solver included in Dr.LiTHO and with the finite-element-method
EMF solver JCMsuite.
We demonstrate results on convergence for dense and isolated hole arrays, for masks including
diagonal structures, and for a large 3D mask pattern of lateral size 10 microns by 10 microns.

\end{abstract}


\section{Introduction}

Support from modeling and simulation is critical to push the limits
of traditional optical lithography. A specific requirement for
lithography modeling and simulation is the need for very efficient
electromagnetic field (EMF) solvers that allow the simulation of
large 3D computational domains~\cite{ITRS2007}. We have developed an
interface between the lithography simulator Dr.LiTHO and the program
package JCMsuite for EMF simulations. Dr.LiTHO is a comprehensive
simulation environment for photolithography developed at the
Fraunhofer IISB. The included EMF
simulators are based on the waveguide
method~\cite{Evanschitzky2007a} and on the FDTD method.
JCMsuite is a finite element based solver for EMF simulations
developed at JCMwave and at the Zuse Institute
Berlin~\cite{Burger2005bacus,Burger2006c,Burger2007bacus}. Both
solvers allow the rigorous simulation of relatively large 3D
computational domains. The interface between the program packages
enables the accuracy benchmarking of the results obtained with the
EMF simulators of Dr.LiTHO and JCMsuite.

We quantitatively compare the near field results
(complex diffraction coefficients) obtained with the
waveguide method and with the finite element method.
Then the resist images resulting from the mask near fields are
computed with the fully vectorial imaging system of Dr.LiTHO.
We also compare the resist images corresponding to the near
field results obtained with the respective EMF solvers regarding
process windows and CDs.

We investigate several application examples:
(a) isolated and dense contact holes with target CD's of 65\,nm and 45\,nm.
(b) Z-like structures in order to get a combination of horizontal,
vertical and 45-degree rotated elements which are critical from
the simulation point of view, and
(c) a larger more complex patterned mask area with a size of
10\,microns x 10\,microns.
For all simulations we consider state of the art mask types and lithography settings.

\section{Background}
\subsection{Dr.LiTHO}
\label{wgm_chapter}

\noindent Dr.LiTHO is a comprehensive simulation environment for
photolithography, developed at the Fraunhofer IISB.\footnote{
URL: \urldrlitho} Its main focus
is on development and research applications. Dr.LiTHO includes
models and algorithms for the simulation, evaluation, and
optimization of lithographic processes using optical or EUV image
projection. One of the models for the rigorous simulation of light
diffraction from lithography masks is the waveguide method.

\subsubsection{Waveguide Method}

\noindent The basic idea of the computation of light diffraction
from a lithography mask with the waveguide method is as follows:
Based on the real mask geometry a slicing of the area to be
simulated is performed by defining maximum sized layers which are
homogeneous in the direction of light propagation (this is the so
called waveguide assumption). The geometry (material distribution)
of the individual layers is described by an arbitrary number of
rectangles with arbitrary sizes (depending on the geometry and on
the required resolution). With this mask description any geometry
can be realized and a potential sampling error can be limited to
the desired
value. Then the material distributions and the electromagnetic
fields of all layers are described by Fourier series. All Fourier
series are truncated according to the number $M$ of modes which are
supposed to be taken into account. The combination of this material
and field descriptions with the Maxwell equations leads to an
eigenvalue problem for each layer and finally to the propagating and
evanescent modes inside the layers. By applying the appropriate
boundary conditions all layers are coupled and the resulting
reflected and transmitted plane waves at the top an bottom side of
the mask are computed. Because of the periodic boundary conditions
in lateral direction (x- y-plane, perpendicular to the direction of
light propagation), the mask is always regarded as a periodic
structure. Isolated features can be simulated by using a mask period
which is large enough for the respective problem. Two dimensional
(e.g. lines) and three dimensional (e.g. contact holes) computations
are possible with the method. Further detailed technical and
mathematical descriptions of the waveguide method including
simulation examples and further information on modal methods can be
found
in~\cite{Evanschitzky2007b,Lucas1996a,Moharam1995a,Turunen1997a}.

\noindent The computation time is proportional to $A\cdot M^{3}$.
$A$ is the number of layers consisting of more than one material
after the described mask slicing. $M$ is the overall number of modes
used for the computation and can be expressed by
$M=(2N_{x}+1)\cdot(2N_{y}+1)$. According to all investigations
performed so far the following is required:
$N_{x,y}=b_{x,y}/2\lambda$ for extreme ultra violet (EUV) masks and
$N_{x,y}=3b_{x,y}/\lambda$ for optical masks ($b_{x,y}$ is the real
mask period in x- and y-direction, and $\lambda$ is the illumination
wavelength).

\noindent In the optical case (e.g. at 193 nm), the convergence is a
basic problem of all electromagnetic field solvers based on modal
methods like waveguide because of the significant refractive index
differences between the materials. An implemented mathematical
optimization reduces this problem effectively with the result of a
very good convergence in this case. In contrast to that the EUV case
is uncritical because the refractive indices of all materials are
close to~$1$.

\noindent If larger mask areas have to be simulated the computation
time of a three dimensional simulation can become too long. A
simplified three dimensional waveguide computation based on a so
called decomposition technique is required. This technique replaces
a full three dimensional simulation by a superposition of several
two dimensional and one dimensional simulations. The result is a
significantly reduced computation time but compared to the
spectrum/near field of a full three dimensional simulation an error
must be accepted. The basic idea of the decomposition technique is
to separate diffraction effects from mask edges along x- and
y-direction (the x- y-plane is perpendicular to the direction of
light propagation) by splitting up a full three dimensional
electromagnetic field simulation into several two dimensional and
one dimensional parts. The application of this method to EUV masks
and "No Hopkins" simulations (rigorous simulation of the off axis
illumination of masks) requires additionally the consistent handling
of off axis illumination which includes also the case of conical
diffraction.

\noindent In the first step the real three dimensional (3D) mask is
split up into two dimensional (2D) parts with homogenous dielectric
properties with respect to the x- and y-direction. In the next step
all 2D parts are computed with the waveguide method under the
assumption that the dielectric properties of the mask vary only in
x- and y-direction. Therefore only 2D simulations are required.
Finally, the transmissions and reflectivities of the mask at all
cross-over areas of the individual 2D parts are computed with the
waveguide method. This is realized by 1D simulations. The resulting
spectrum of the three dimensional mask is obtained by a composition
of the complex spectra of the x- and y-configurations and the mask
transmissions and reflectivities. For a better performance and
accuracy the composition of the partial results is performed in the
frequency domain. Simulation examples of waveguide with
decomposition technique and comparisons with full 3D simulations can
also be found in previous works~\cite{Erdmann2003a,Evanschitzky2007a}.

\begin{figure}[htb]
\centering
\psfrag{Userinput}{\sffamily User Input}
\psfrag{EMF}{\sffamily EMF simulation}
\psfrag{Imaging}{\sffamily Imaging simulation}
\psfrag{Resist}{\sffamily Resist simulation}
\psfrag{maskgeo}{\sffamily Mask geometry}
\psfrag{materials}{\sffamily Material parameters}
\psfrag{illumination}{\sffamily Illumination param.}
\psfrag{imaging}{\sffamily Imaging parameters}
\psfrag{resist}{\sffamily Resist parameters}
\psfrag{nearfieldfem1}{\sffamily JCMsuite (FEM)}
\psfrag{nearfieldfem2}{\sffamily e-m near field}
\psfrag{nearfieldwgm1}{\sffamily Dr.LiTHO (WGM)}
\psfrag{nearfieldwgm2}{\sffamily e-m near field}
\psfrag{aerialimage}{\sffamily Aerial image}
\psfrag{resistimage}{\sffamily Resist image}
\psfrag{resistpattern}{\sffamily Resist pattern}
\psfrag{}{\sffamily }
\fbox{\hspace{0.5cm} \includegraphics[width=0.9\textwidth]{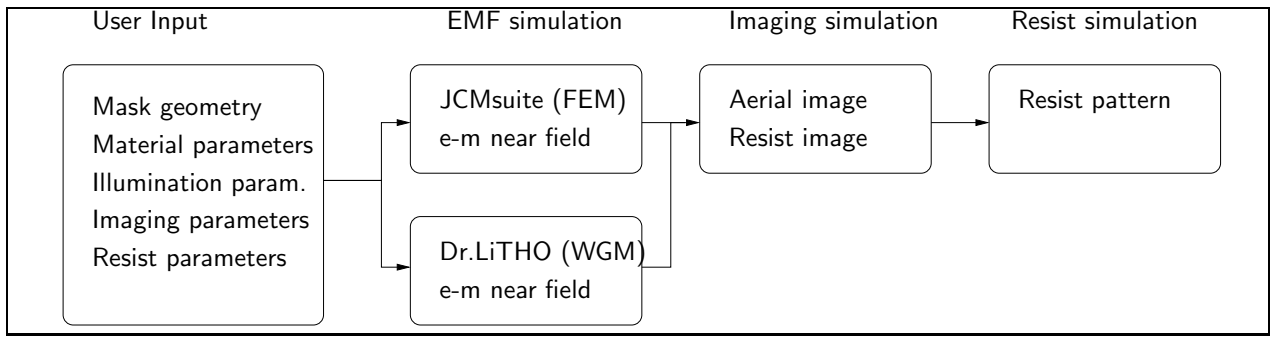}\hspace{0.5cm}}
\caption{
Schematics of the interface between Dr.LiTHO and JCMsuite.
}
\label{schema_interface}
\end{figure}
\subsection{JCMsuite}
JCMsuite is a finite-element package for accurate and fast simulations of
electromagnetic problems.\footnote{URL: \urljcmwave} Due to the good convergence properties of FEM
it is especially well suited for the accurate simulation of nanostructures,
e.g., in photomask simulations, optical metrology and integrated optics.
Comparably large 3D computational domains can be handled at moderate computational
effort~\cite{Burger2006c,Burger2007bacus}.
The solver has been compared and benchmarked with RCWA and FDTD-based EMF simulators
for 2D~\cite{Burger2005bacus} and 3D~\cite{Burger2006c} computational domain problems.

\begin{figure}[htb]
\centering
\psfrag{CDx}{\sffamily $\mbox{CD}_{\mbox{x}}$}
\psfrag{CDy}{\sffamily $\mbox{CD}_{\mbox{y}}$}
\psfrag{px}{\sffamily $\mbox{p}_{\mbox{x}}$}
\psfrag{py}{\sffamily $\mbox{p}_{\mbox{y}}$}
\fbox{\hspace{0.5cm} \includegraphics[height=0.3\textwidth]{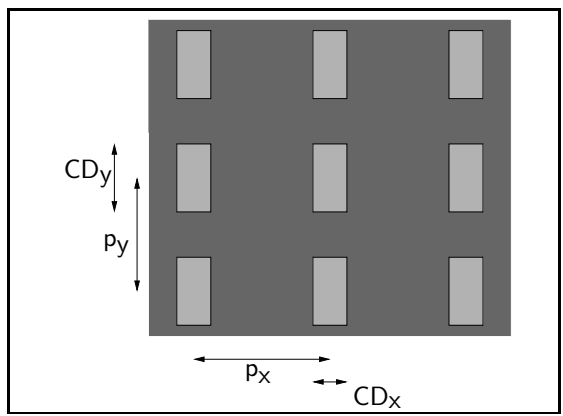}\hspace{0.5cm}}
\caption{
Schematics of a 2D-periodic array of contact holes: cross-section in a $x$-$y$-plane.
Holes with critical dimensions of $CD_x$, $CD_y$  are depicted
in light grey. The pattern is periodic in $x/y$-direction with a pitch of $p_x/p_y$.
}
\label{schema_geo_contact_hole}
\end{figure}

\subsubsection{Finite Element Method}

Light propagation in 3D photomasks is governed by Maxwell's
equations.
The finite-element method is used to find rigorous solutions to these. The
method consists of the following steps:
\begin{itemize}
\item
The geometry of the computational domain is discretized with simple geometrical patches,
{\it JCMsuite} uses tetrahedral or prismatoidal (3D) patches.
The use of prismatoidal patches is an advantage for layered geometries, as in
photomask simulations.
Sidewall angles different from 90\,deg are not regarded throughout this paper;
however, they are easily be implemented and do not lead to additional computational effort.
Geometries consisting of periodic arrangements and of isolated patterns are both
treated rigorously. The computational effort for isolated problems is substantially
decreased because a large-pitch, quasi-periodic model is not
required~\cite{Burger2007bacus,Zschiedrich2008pmj}.
\item
The function spaces in the integral representation of Maxwell's equations
are discretized using Nedelec's edge elements,
which are vectorial functions of polynomial order
defined on the simple geometrical patches~\cite{Monk2003a}.
In the current implementation, {\it JCMsuite} uses polynomials of
first ($1^{st}$) to ninth ($9^{th}$) order.
In a nutshell, FEM can be explained as expanding the field
corresponding to the exact solution of Maxwell's equations in the
basis given by these elements.
\item
This expansion leads to a large sparse matrix equation (algebraic problem).
To solve the algebraic problem on a standard workstation
linear algebra decomposition techniques (e.g., sparse LU-factorization)
are used.
In cases with either large computational domains or high accuracy
demands, also rigorous domain decomposition methods~\cite{Zschiedrich2008al}
are used and allow to handle problems with very large numbers of unknowns.
\end{itemize}

For details on the weak formulation,
the choice of Bloch-periodic functional spaces,
the FEM discretization, and the implementation of the adaptive PML
method in {\it JCMsuite}
we refer to previous works~\cite{Pomplun2007pssb}.

\begin{table}[h]
\begin{center}
\begin{tabular}{|l||l|l|l||l|l|}
\hline
 & Layout 1.1 & Layout 1.2 & Layout 1.3 & Layout 2.1 & Layout 2.2 \\
\hline
\hline
$\mbox{CD}_{\mbox{x}}$ [nm]  & 240 & 180 & 180 & 255  & 370 \\
$\mbox{CD}_{\mbox{y}}$ [nm]  & 300 & 240 & 240  & 180  & 260 \\
$\mbox{L}_{\mbox{x}}$ [nm]  &&& & 900  & 1300 \\
$\mbox{L}_{\mbox{y}}$ [nm]  &&& & 1000 & 1450  \\
$p_x$ [nm]                       & 480 & 360 & 3600 & 1500 & 2000\\
$p_y$ [nm]                       & 600 & 480 & 4800 & 1500 & 2000 \\
\hline
material stack              & Mat2 & Mat1 & Mat1 & Mat1 & Mat2 \\
\hline
\end{tabular}
\caption{Parameter settings for the contact hole
mask simulations (Layout 1.1 -- 1.3, compare Fig.~\ref{schema_geo_contact_hole})
and for simulations of
$Z$-like structures (Layout 2.1 -- 2.2, compare Fig.~\ref{schema_geo_z}):
Mask geometry, material parameters, illumination, and imaging parameters.
}
\label{table_layout_definitions}
\end{center}
\end{table}

\begin{table}[h]
\begin{center}
\begin{tabular}{|l|l|l|l||l|l||l|l|}
\hline
Material stack & & Mat1 & Mat2 & \multicolumn{2}{|l||}{Illumination}& \multicolumn{2}{|l|}{Imaging system} \\
\hline
\hline
$\varepsilon_{r \mbox{air}}$ & 1.0 &\multicolumn{2}{l||} {$\inf$}&$\lambda_0$ & 193\,nm&Magnification & $4\,X$\\
\hline
$\varepsilon_{r \mbox{Cr}}$ & $(0.861 + 1.668i)^2$ & & &Polarization & unpolarized&NA & 1.35 \\
$\mbox{d}_{\mbox{Cr,top}}$ [nm] && 80.0 & -- & & & & \\
\hline
$\varepsilon_{r \mbox{MoSi}}$ & $(2.442 + 0.586i)^2$ & & &Annular & 0.75 - 0.95&Immersion: $n$ & 1.44\\
$\mbox{d}_{\mbox{MoSi}}$ [nm] & & -- & 68.0  & & & &  \\
\hline
$\varepsilon_{r \mbox{SiO}_2}$ & $(1.5595)^2$ & \multicolumn{2}{l||} {$\inf$}& \multicolumn{4}{l|} {}\\
\hline
\end{tabular}
\caption{Parameter settings for the used material stacks: relative permittivities and layer thicknesses
are specified for the different investigated mask stacks. Parameters of the illumination and fully vectorial imaging system simulations for all simulation
results presented in this paper.
}
\label{table_material_stacks}
\end{center}
\end{table}

\section{Benchmark of the Waveguide Method and the Finite Element Method}
\label{benchmarkchapter}

We have developed an interface between the Lithography simulator
Dr.LiTHO and the EMF simulator JCMsuite. As will be shown in the
quantitative comparisons this combination of advanced lithography
and aerial image simulation (Dr.LiTHO) with a advanced
electromagnetic field simulator (JCMsuite) allows to treat demanding
lithography problems fast and accurately.
Figure~\ref{schema_interface} shows a schematics of the simulation
flow: The user defines mask geometry, material parameters,
illumination and imaging parameters and resist parameters. Then
either the EMF simulator JCMsuite computes the electromagnetic near
field distribution. Alternatively the near field distribution can be
computed using Dr.LiTHO's field simulator based on WGM. This allows
for a quantitative comparison of the achieved results. From the near
field results of either method then Dr.LiTHO's aerial image
simulator generates the aerial image or resist image. And from this
by using Dr.LiTHO's resist simulator the resist topography
after development is generated~\cite{Fuehner2007al}.
For demonstrating the accuracy and capabilities of this combination we benchmark near field and
far field results obtained with the two different near field simulators for several application
problems.

\subsection{Isolated and dense contact holes}
\label{contact_hole_chapter}

Figure~\ref{schema_geo_contact_hole} shows a schematics of the simulated periodic arrays of
square-shaped holes. We investigate three cases:
\begin{itemize}
\item Layout 1.1: An attenuated phase-shift mask (MoSi) with contact holes for a target CD of 65\,nm.
      The pattern is dense (CD\,:\,Pitch = 1\,:\,2).
\item Layout 1.2: A chromium mask with contact holes for a target CD of 45\,nm.
      The pattern is dense (CD\,:\,Pitch = 1\,:\,2).
\item Layout 1.3: A chromium mask with contact holes for a target CD of 45\,nm.
      The pattern is isolated (CD\,:\,Pitch = 1\,:\,20).
\end{itemize}

All geometry parameters and the used material stacks, illumination and imaging parameters
are listed in Tables~\ref{table_layout_definitions} and~\ref{table_material_stacks}.
Tables~\ref{table_contact_hole_results} and~\ref{table_contact_hole_results_1_3} show
results of the benchmark and convergence investigations:
We perform simulations using the FEM near field simulator JCMsuite and using the WGM solver Dr.LiTHO.
For both solvers we use different numerical settings in order to demonstrate convergence of the results.
The varied numerical parameter for the FEM solver is the polynomial degree $p$ of the used finite-element
ansatz-functions. For WGM the parameter $p$ is the truncation number of the Fourier basis.
We further list computation times and (for FEM) numbers of unknowns of the discrete problem.
From the simulated near fields and spectra we compute the aerial images and the process windows using
Dr.LiTHO's aerial imaging simulator.
From the aerial images, we compute the critical dimensions at a fixed threshold, $CD_x$ and $CD_y$.
This threshold is choosen such that $CD_x$ is on target for the highest numerical resolution.
Similarly, the computed process windows are centered around a fixed threshold obtained from the
simulation with highest numerical resolution.
To demonstrate convergence of the near field results we further list the intensities of two distinct
diffraction orders in the scattering matrix,
$I(0,0)$, $I(1,-1)$.
The complex valued scattering matrix is the input to Dr.LiTHO's aerial imaging simulation.

\begin{figure}[htb]
\centering
\fbox{\hspace{0.5cm}
\includegraphics[width=0.4\textwidth]{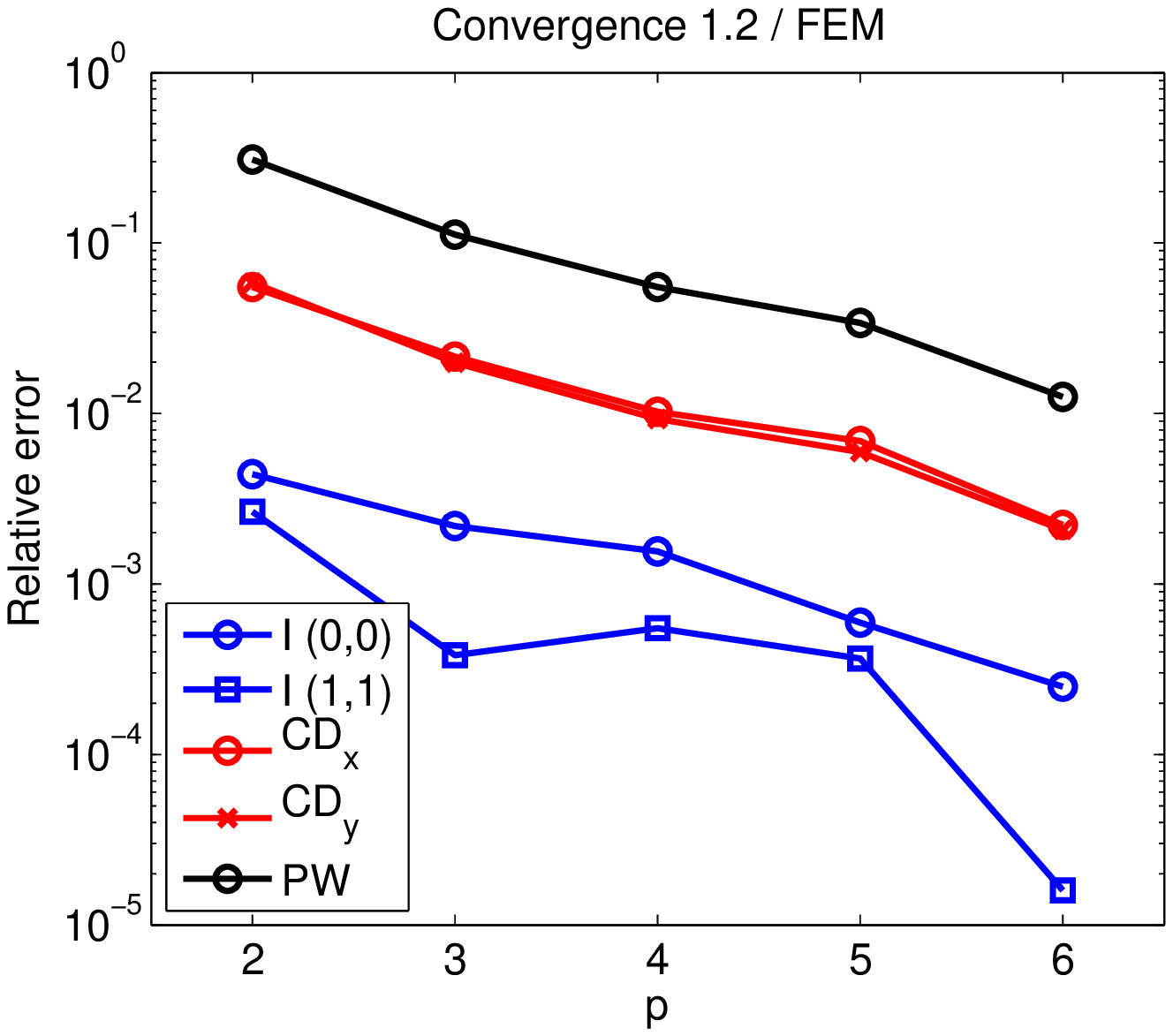}
\hspace{0.5cm}}
\caption{
Convergence of the FEM results for Layout~1.2.
Blue markers show the relative error of the near field results in dependence on the parameter
$p$, red markers show the relative error of the aerial image CD's at fixed threshold, and
black markers show the relative error of the corresponding process window width.
(Compare Table~\ref{table_contact_hole_results}.)
 }
\label{conv_1}
\end{figure}

\begin{table}[h]
\begin{center}
\begin{tabular}{|l|r|r|r|l|l|l|l|l|}
\hline
\multicolumn{9}{|l|}{Layout 1.1}\\
\hline
 & p & t [s] & N & $I(0,0)$ & $I(1,-1)$ & $CD_x$ [nm]& $CD_y$ [nm]& $\Delta PW$ [nm]\\
\hline
 JCMsuite
(FEM)&2 & 2 &11016    & 0.037864 & 0.174428 &  64.68 & 70.43 &  53.52 \\
&3 & 4 &13160    & 0.044147 & 0.172952 &  64.07 & 69.83 &  47.49 \\
&4 & 16 &28550   & 0.045656 & 0.173134 &  64.67 & 70.71 &  56.27 \\
&5 & 42&42192    & 0.045554 & 0.173092 &  64.84 & 70.63 &  57.61 \\
&6 & 102&61397   & 0.045387 & 0.173101 &  64.82 & 70.65 &  57.46 \\
&7& 270&108352   & 0.045557 & 0.173121 &  64.99 & 70.86 &  59.57 \\
&8& 578&158328   & 0.045585 & 0.173120 &  65.00 & 70.86 &  59.64 \\
\hline
 Dr.LiTHO
 (WGM)&    1 &  &   & 0.160890   & 0.204421 & & & \\
 &    2 &  &   & 0.127415   & 0.174654 & & & \\
 &    3 &  &   & 0.090717   & 0.177855 & & & \\
 &    4 &  &   & 0.072831   & 0.171335 & & & \\
 &    5 &  &   & 0.058509   & 0.175867 & & & \\
 &    6 &  &   & 0.057931   & 0.173554 & & & \\
 &    7 &  &   & 0.052896   & 0.173336 & & & \\
 &    8 &  &   & 0.050502   & 0.173694 & & & \\
 &    9 &  &   & 0.050156   & 0.173185 & & & \\
 &   10 & 74 & & 0.048607   & 0.173374 &  66.94 & 72.86  & 79.64\\
\hline
\hline
\multicolumn{9}{|l|}{Layout 1.2}\\
\hline
 & p & t [s] & N & $I(0,0)$ & $I(1,-1)$ & $CD_x$ [nm]& $CD_y$ [nm]& $\Delta PW$ [nm]\\
\hline
 JCMsuite
(FEM) &2 & 1 & 7803   & 0.306478 & 0.126153 & 47.49 & 40.91  & 121.00 \\
 &3 & 4 & 11844  & 0.305795 & 0.125770 & 45.97 & 39.41  & 102.81 \\
 &4 & 16 & 22840 & 0.305602 & 0.125749 & 45.46 & 39.00  & 97.58 \\
 &5 & 40 & 42192 & 0.305311 & 0.125864 & 45.31 & 38.87  & 95.62 \\
 &6 & 114 & 70168& 0.305206 & 0.125816 & 45.10 & 38.72  & 93.63 \\
 &7 & 264 &108352& 0.305130 & 0.125818 & 45.00 & 38.64  & 92.47  \\
\hline
Dr.LiTHO
 (WGM)&    2 &  &   & 0.294604 & 0.115876 & 26.30 & 23.37  & NaN \\
 &    4 &  &   & 0.304413 & 0.124390 & 43.33 & 38.07  &  86.17  \\
 &    6 &  &   & 0.303733 & 0.125069 & 42.60 & 37.21  &  72.21  \\
 &    8 & 12 &   & 0.303683 & 0.125334 & 42.57 & 37.15  &71.15  \\
\hline

\end{tabular}
\caption{Convergence of near field and far field results for the different examples
of Chapter~\ref{contact_hole_chapter} and for the different methods.
The table shows some numerical results for different accuracy settings of the respective software.
For JCMsuite, $p$ corresponds to the polynomial degree setting of the finite-element
ansatz-functions, $t$ corresponds to the CPU time in seconds,
and $N$ corresponds to the number of unknowns in the FEM problem.
For $p\leq 4$, JCMsuite computations have been performed using a single thread, for $p>4$, eight threads
of a standard multi-core workstation
have been used.
For Dr.LiTHO, $(2p+1)^2$ is the number of Fourier modes in the Waveguide method.
$I(0,0)$ and $I(1,-1)$ are the sums of the magnitudes of the entries in the
scattering matrix for the zero$^{th}$ and for the $(1,-1)$ diffraction order for
TE and TM polarization.
Dr.LiTHO computations have been performed using a single thread of a standard personal computer.
$CD_x$ and $CD_y$ are aerial image CDs at fixed threshold (choosen such that $CD_x$ at highest FEM
resolution is on target).
$\Delta PW$ is the width of a rectangular process window
with a fixed exposure $\pm 2.5\%$
in a plot of exposure versus defocus for $CD_x$ contours of the nominal $CD_x$ $\pm 10\%$.
}
\label{table_contact_hole_results}
\end{center}
\end{table}

\begin{table}[h]
\begin{center}
\begin{tabular}{|l|r|r|r|l|l|l|l|l|l|}
\hline
\multicolumn{10}{|l|}{Layout 1.3}\\
\hline
 & p & t [s]& N      & RAM  & $I(0,0)$ & $I(1,-1)$ & $CD_x$ & $CD_y$ & $\Delta PW$ \\
&    &          &   & [GB] & &                    &  [nm]&  [nm]    & [nm]\\
\hline
JCMsuite (FEM)
&2  & 4 &11373   & 0.05& 0.0265226 & 0.0028276 & 46.38 & 48.37  & 56.15 \\
&3 & 10&17244    & 0.15& 0.0266462 & 0.0028365 & 46.30 & 48.32  & 56.09  \\
&4 & 40&33240    & 0.45& 0.0266676 & 0.0028324 &  46.32 & 48.20  &  55.29 \\
&5 & 62 &61392   & 1.0& 0.0264626 & 0.0028299 &  46.12 & 47.90  &  53.85 \\
&6 & 170 &102088 & 2.1& 0.0265100 & 0.0028223 & 45.36 & 47.22  &  50.45 \\
&7 & 412 &157632 & 3.9& 0.0265003 & 0.0028226 & 45.00 & 46.93  &  49.01 \\
\hline
Dr.LiTHO (WGM)
 &    5 &  &   & & 0.0185761 & 0.0051559 & & & \\
 &   10 &  &   & & 0.0240660 & 0.0018821 & & & \\
 &   15 &  &   & & 0.0260278 & 0.0023286 & & & \\
 &   20 &  &   & & 0.0256536 & 0.0025602 & & & \\
 &   25 &  &   & & 0.0255844 & 0.0031166 & & & \\
 &   30 &  &   & & 0.0261552 & 0.0030079 & & & \\
 &   35 &  &   & & 0.0262580 & 0.0028048 & & & \\
 &   40 &  &   & & 0.0262528 & 0.0028277 & & & \\
 &   45 &  &   & & 0.0262878 & 0.0029080 & & & \\
 &   50 &  &   & & 0.0263020 & 0.0028480 & & & \\
 &   55 &  &   & & 0.0262804 & 0.0027809 & & & \\
 &   60 &  &   & & 0.0262583 & 0.0027705 & & & \\
 &   65 &  &   & & 0.0262644 & 0.0027904 & & & \\
 &   70 &  &   & & 0.0262812 & 0.0027847 & & & \\
 &   75 & 6 &  & 0.01 & 0.0262767 & 0.0027663 & 49.77 & 52.11  & 72.42\\
\hline

\end{tabular}
\caption{Convergence results for Layout~1.3, description see Table~\ref{table_contact_hole_results}.
RAM denotes the approximate memory consumption in GB.
WGM uses the decomposition strategy.}
\label{table_contact_hole_results_1_3}
\end{center}
\end{table}

Table~\ref{table_contact_hole_results} lists results for the cases
of dense arrays of contact holes. For the attenuated phase shift
mask, Layout~1.1, using FEM we observe convergence of the near field
intensities, $I(0,0)$, $I(1,-1)$, to a level of about $10^{-5}$. As
can be seen from the table this corresponds to an accuracy of the
aerial image CD's of about 0.01\,nm and an accuracy of the process
window of about 0.1\,nm defocus. Computation times vary between two
seconds and 10 minutes, depending on accuracy setting. Using the
waveguide method for the same example, convergence of the near field
coefficients can also clearly be observed. However, at the highest
used numerical setting of an accuracy level of about  $10^{-3}$ is
reached. This results in a
difference between the methods
of about 2\,nm in aerial image CD's and of about 20\,nm in the
process window (cf., Table~\ref{table_contact_hole_results}). Please
note that the patterns investigated throughout this paper are not
optimized for a maximum process window. Process window errors are
expected to be smaller with aerial image error for optimized process
windows. Computation time for the highest accuracy setting of WGM is
74\,s. 
%

For the chromium mask, Layout~1.2, we observe a similar behavior.
While the highest used FEM accuracy setting reaches an accuracy of smaller 0.1\,nm in CD and 1\,nm in
process window at moderate computational effort (about five minutes computation time on a multi-core
workstation), the highest used WGM accuracy reaches an accuracy of about 2\,nm in CD and 20\,nm in process
window at low computational effort (few seconds computation time on a standard PC).

It is interesting to note that even when the differences of the investigated diffraction
order intensities are only relatively small (e.g., differences only in the third significant digit)
there are
significant differences in the resulting aerial images and in the process windows (e.g., differences
in the first or second significant digit).
Figure~\ref{conv_1} shows the convergence behavior of the
FEM near field and far field results graphically.

Table~\ref{table_contact_hole_results_1_3} lists results for the case of a periodic array of contact holes
with large pitch (quasi isolated case), Layout~1.3.
Due to the large size of the computational domain, in this case the WGM solver is switched to
the decomposition mode (see Chapter~\ref{wgm_chapter}).
This approach contains some approximations to the rigorous model, therefore it is expected that the results
do not exactly converge to the numbers obtained with the finite element method.
As can be seen from the table, the higher order near field diffraction orders converge to numbers with a few percent
offset from the rigorous results.
The CD's obtained with FEM and obtained with WGM differ by about 5\,nm and the
process windows differ by about 20\,nm.

\subsection{Z-like structures}
\label{z_like_subsection}
\begin{figure}[htb]
\centering
\psfrag{CDx}{\tiny \sffamily $\mbox{CD}_{\mbox{x}}$}
\psfrag{CDy}{\sffamily \small $\mbox{CD}_{\mbox{y}}$}
\psfrag{px}{\sffamily $\mbox{p}_{\mbox{x}}$}
\psfrag{py}{\sffamily $\mbox{p}_{\mbox{y}}$}
\psfrag{Lx}{\sffamily $\mbox{L}_{\mbox{x}}$}
\psfrag{Ly}{\sffamily $\mbox{L}_{\mbox{y}}$}
\fbox{\hspace{0.5cm}
{ \sffamily a) }
\includegraphics[height=0.3\textwidth]{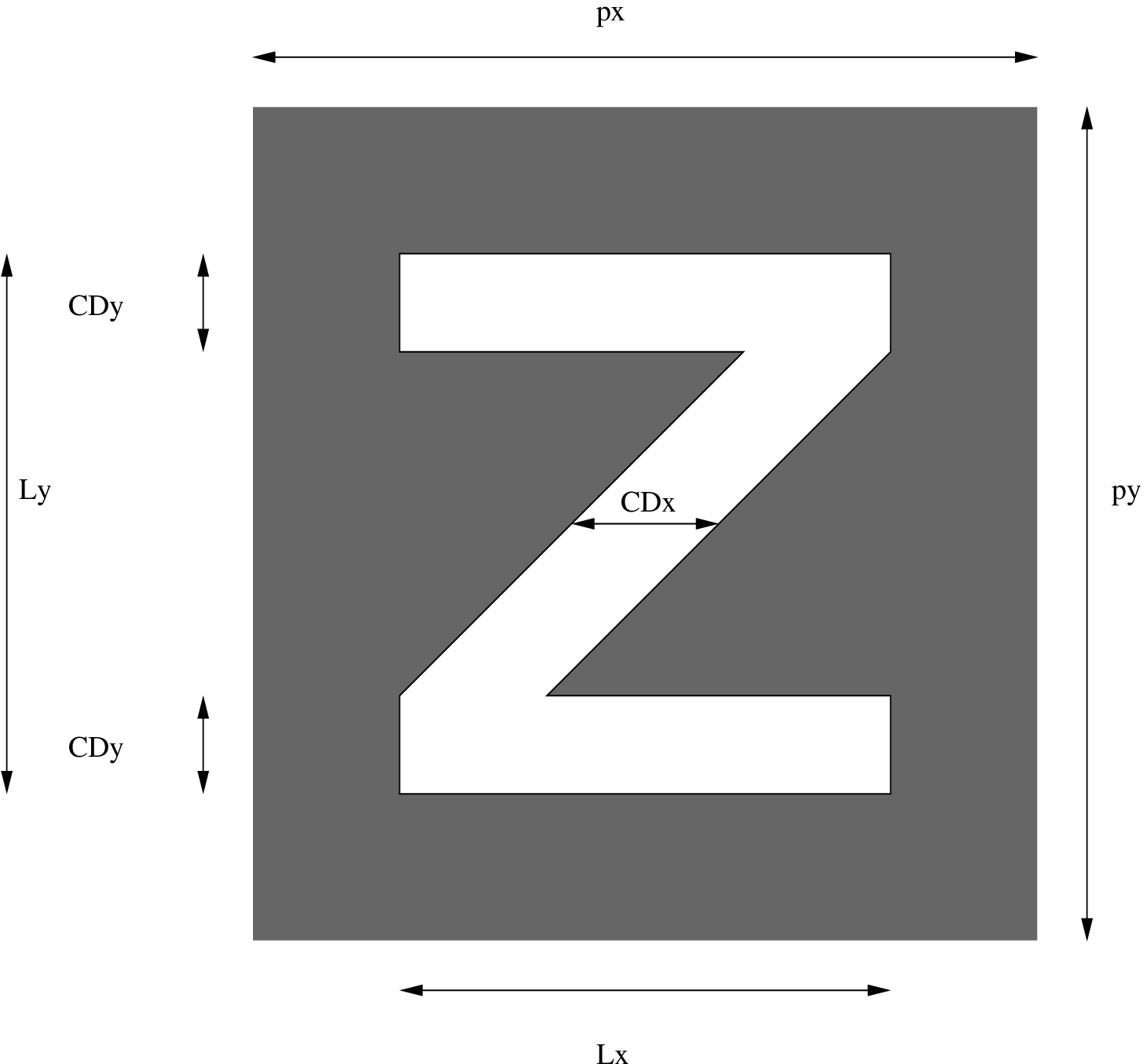}
{ \sffamily b) }
\includegraphics[height=0.3\textwidth]{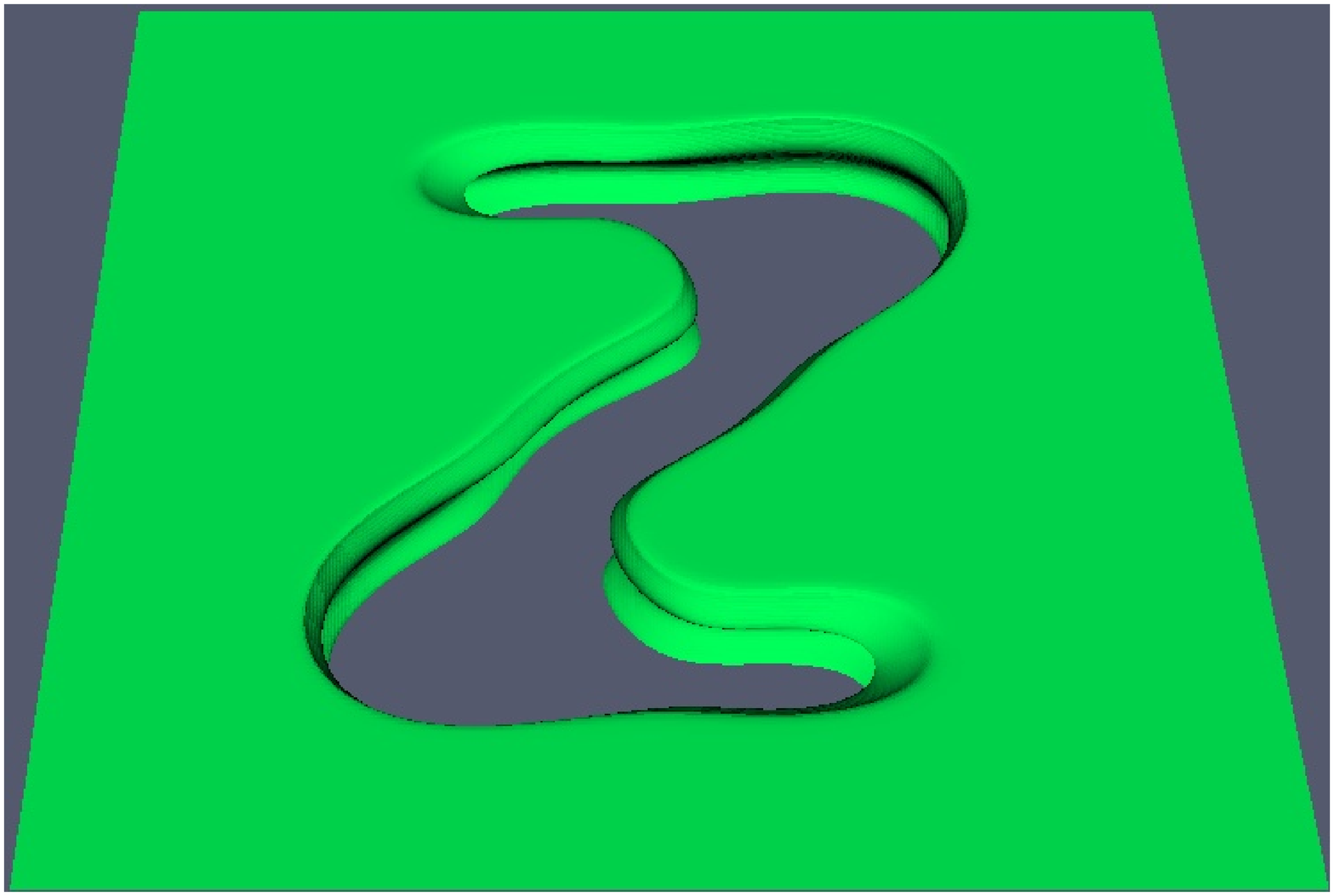}
\hspace{0.5cm}}
\caption{
a) Schematics of the investigated structures with a 45-degree angle (Z-like structures):
cross-section in a $x$-$y$-plane.
Structures with critical dimensions of $CD_{y}$, $CD_y$ and total dimensions $L_x$, $L_y$, are depicted
in white. The pattern is periodic in $x/y$-direction with a pitch of $p_x/p_y$.
b) Resist profile: Result from a lithography simulation using the combination of JCMsuite's electromagnetic field solver and
Dr.LiTHO's aerial imaging and resist simulation modules.
}
\label{schema_geo_z}
\end{figure}

In this section we investigate structures which include 45-degree angles in the $x$-$y$-plane.
These structures are favourable from the electronic design point of view, however,
they are critical for optical simulators relying on $x$-$y$-structured meshes.
Figure~\ref{schema_geo_z} shows a schematics of the Z-like structures.
We investigate two cases for structures with minimal
critical dimensions of 45\,nm, resp. 65\,nm ($1\,X$).
The corresponding geometrical parameters are listed in Table~\ref{table_layout_definitions}
(Layout~2.1, 2.2).

\begin{table}[h]
\begin{center}
\begin{tabular}{|l|r|r|r|r|l|l|l|l|l|}
\hline
\multicolumn{10}{|l|}{Layout 2.1}\\
\hline
 & p & t [s]& N      & RAM  & $I(0,0)$ & $I(1,-1)$ & $CD_x$ & $CD_y$ & $\Delta PW$ \\
&    &          &   & [GB] & &                    &  [nm]&  [nm]    & [nm]\\
\hline
 JCMsuite (FEM)
 &2 & 6 & 33810   &0.3 & 0.29313 & 0.07655  & 65.12 & 40.22  &  200.13 \\
 &3 & 23 & 50400  &0.7 & 0.29127 & 0.07750  & 66.07 & 38.58  &  197.27 \\
 &4 & 53 & 63700  &1.1 & 0.29001 & 0.07652  & 60.93 & 38.18  &  174.42 \\
 &5 & 73 & 84000  &2.0 & 0.29100 & 0.07720  & 62.57 & 39.25  &  183.98 \\
 &6 & 143 & 111720&3.1 & 0.29133 & 0.07743  & 63.81 & 39.00  &  188.74 \\
 &7 & 369 & 172480&5.9 & 0.29084 & 0.07751  & 63.75 & 38.32  &  186.88 \\
\hline
 Dr.LiTHO (WGM)
 &  23   & 36400 & &1.7   &  &  &  52.25 & 35.66  & 92.46 \\
\hline
\hline
\multicolumn{10}{|l|}{Layout 2.2}\\
\hline
 JCMsuite (FEM)
 &2 & 17 & 63798   &0.7& 0.05471 & 0.10008 &  82.73 & 52.88  & NaN \\
 &3 & 54 & 82320   &1.3& 0.05544 & 0.09985 &  87.30 & 57.06  & 58.23  \\
 &4 & 172 & 127400 &2.7& 0.06267 & 0.09916 &  89.14 & 60.81  &  90.02  \\
 &5 & 196 & 188160 &4.9& 0.06583 & 0.10022 &  94.36 & 62.36  &  108.12 \\
 &6 & 227 & 156408 &4.4& 0.06428 & 0.10157  &  94.24 & 61.48  &  105.10  \\
 &7 & 509 & 241472 &8.0& 0.06302 & 0.10156  &  92.50 & 60.94  &   94.83  \\
\hline
Dr.LiTHO (WGM)
 &     & 182  &   &0.1 & &  &  86.01 & 65.04  &  31.10\\
\hline
\end{tabular}
\caption{Convergence of near field and far field results for the different examples
of Chapter~\ref{z_like_subsection} and for the different methods.
$CD_x$ and $CD_y$ are aerial image CDs at fixed threshold (choosen such that $CD_x$ at highest FEM
resolution is on target).
$\Delta PW$ is the width of a rectangular process window
with a fixed exposure $\pm 2.0\%$
in a plot of exposure versus defocus for $CD_x$ contours of the nominal $CD_x$ $\pm 10\%$.
For Layout~2.1, WGM has been used in the full-3D (rigorous) mode, for Layout~2.2, WGM has
been used in decomposition mode.
}
\label{table_z_results}
\end{center}
\end{table}

Table~\ref{table_z_results} shows convergence results for the
investigated cases of Z-like structures. Due to the larger
computational domain and the more complex structure, the
computational effort is larger than in the contact hole cases,
Chapter~\ref{contact_hole_chapter}. In the first case (Chromium
material stack, Layout~2.1), FEM shows convergence of the CD's with
an accuracy of about 0.1-1\,nm, and of the process window of about
2\,nm. Computation time for the most accurate FEM result was about
5\,minutes (using 8~cores of a multi-processor workstation). The
full 3D waveguide method result with a trunctation number of 23
yields results with a CD
difference between the methods
in the 10\%-range, and consequently also a large numerical
difference between
the process window width results. Due to the high number of Fourier
modes taken into account, the computation time was several hours
(using a single processor on a standard PC).

Probably due to the more involved material properties of the second example of this chapter, Layout~2.2,
the convergence is less pronounced than in the first example.
For FEM, CD accuracies in the range of 2\,nm are reached at computation times below
10~min, and the process window width seems still not to be clearly converged to a certain region.
The waveguide method (in decomposition mode) yields CD results which are different from the FEM results
by about 5\,nm (with different signs for $x$- and $y$-directions)
and a process window width which is about one third of the FEM process window width.

The case of a geometry with a 45-degree angle is clearly more straight-forward for a FEM solver,
because FEM does not rely on regular grids and can resolve arbitrary angles and shapes.
Therefore the advantage of FEM in the comparison of the rigorous solvers (for Layout~2.1) is
clear.
The case is also involved for the decomposition strategy of the waveguide method, due to the mutual influence of
the various decomposed regions which is partly neglected in this strategy.

Figure~\ref{schema_geo_z}\,b) shows the resist topography after etch. This resist simulation
has been performed using a 80\,nm resist and further undisclosed resist parameters
without any optimization.
This simulation result demonstrates the availability of a full lithography simulation environment for
highly accurate simulations including advanced geometrical mask setups.

\subsection{Larger and more complex patterned mask}
\label{larger_mask_subsection}
\begin{figure}[htb]
\centering
{\sffamily a)
\includegraphics[height=0.23\textwidth]{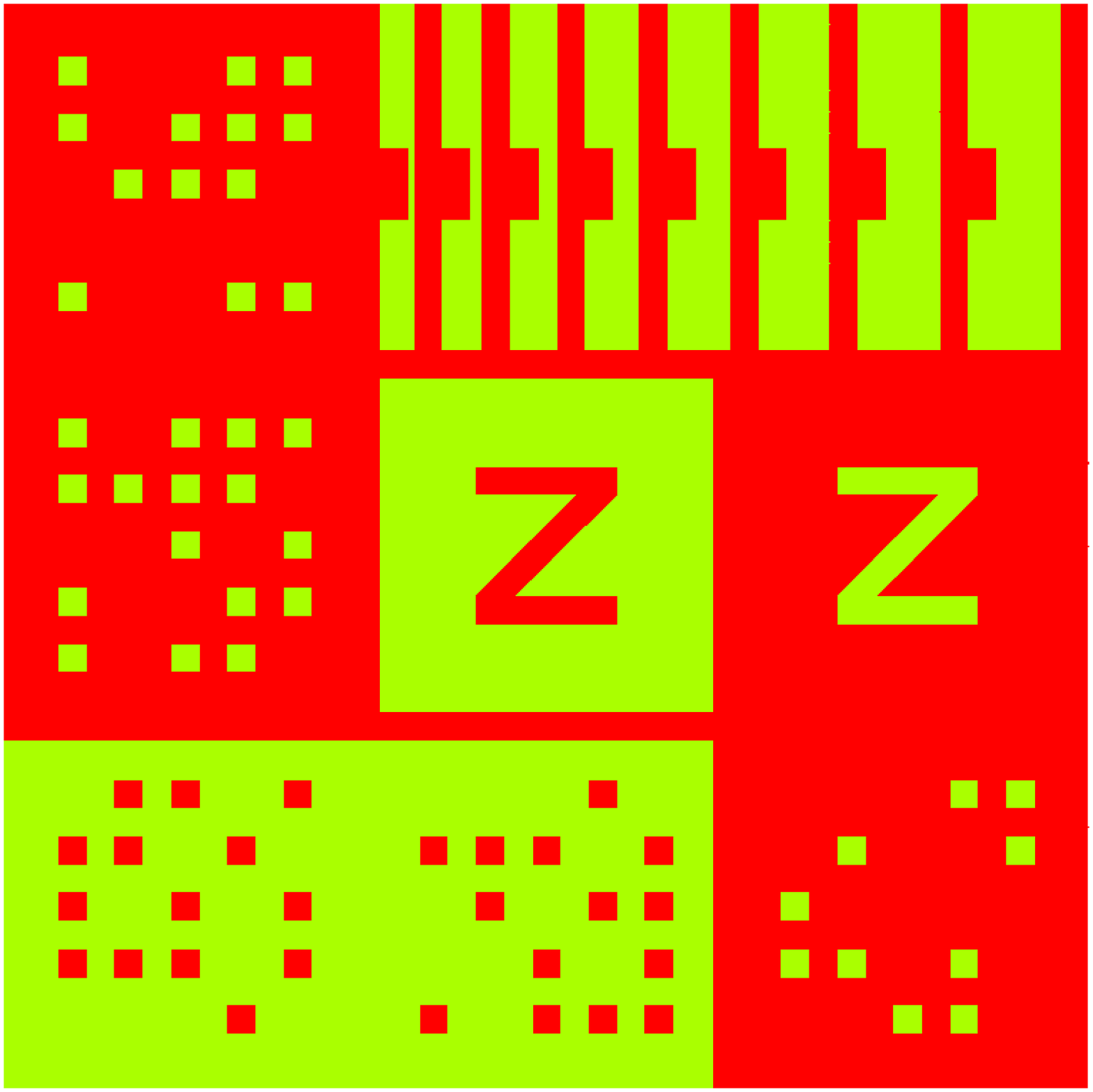}
b)
\includegraphics[height=0.23\textwidth]{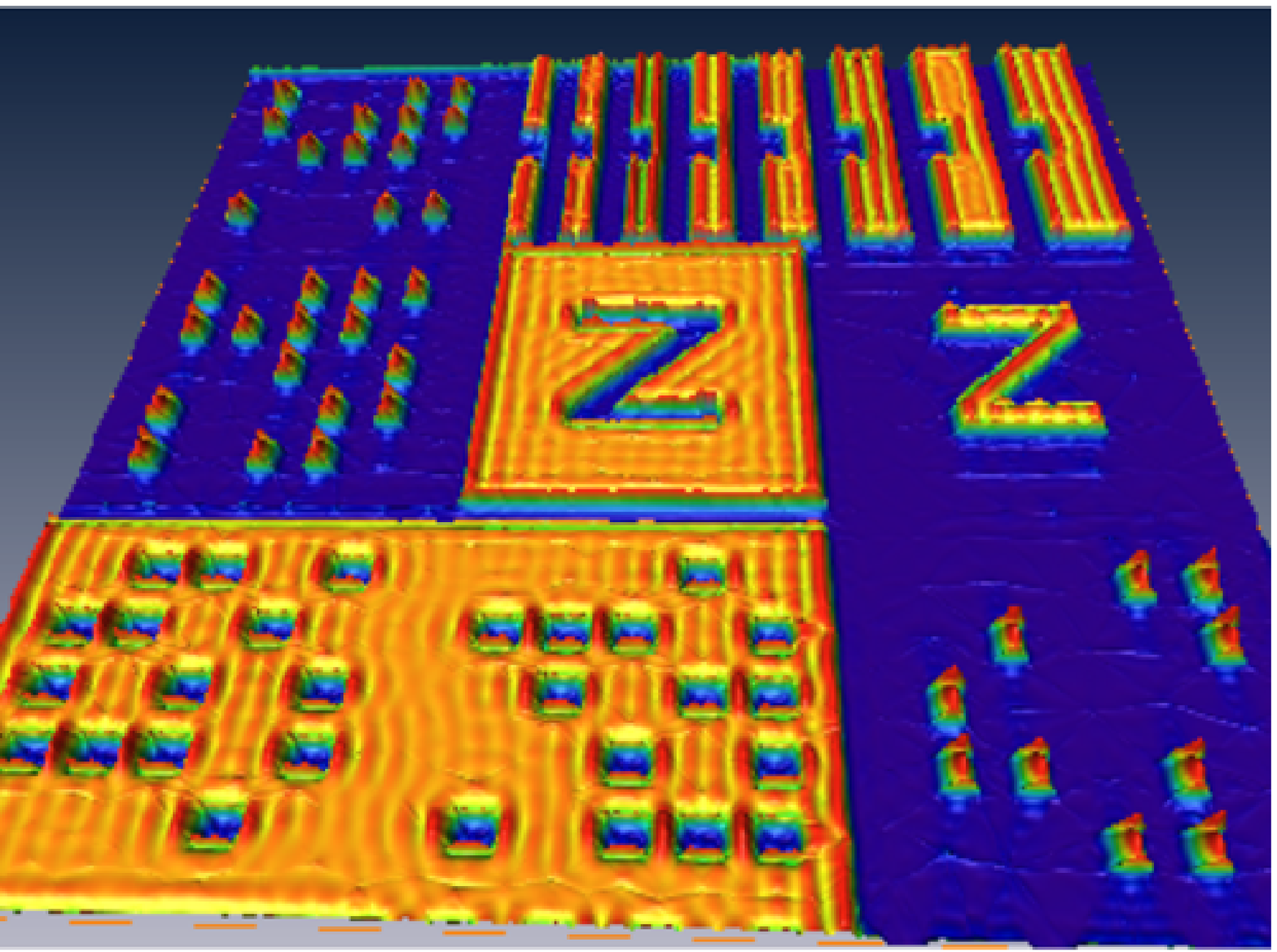}
c)
\includegraphics[height=0.23\textwidth]{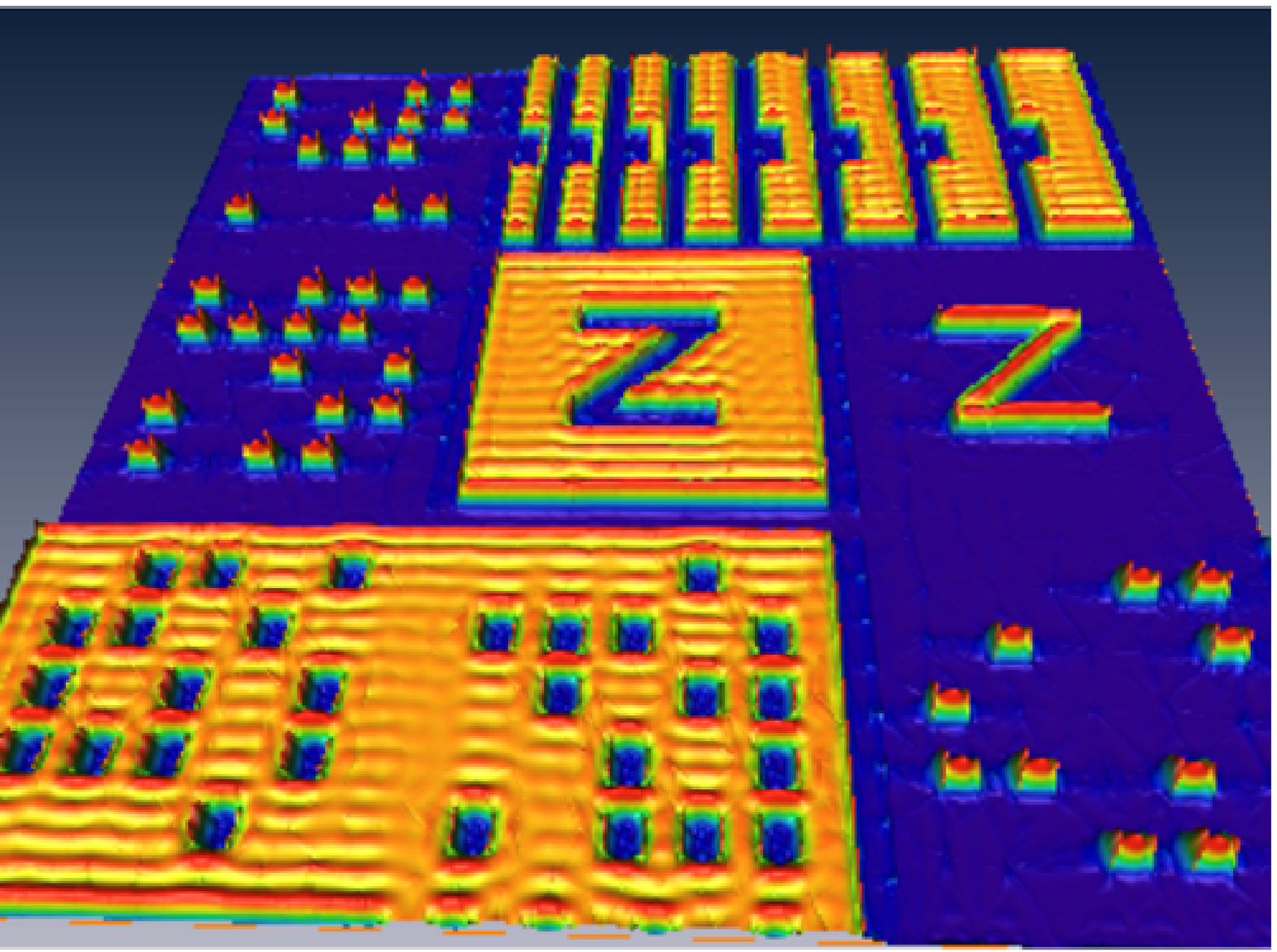}
}
\caption{
a) Schematics of the larger mask (10microns by 10 microns).
b,c) Near field intensity distribution in a pseudo-color visualization for two incident field
polarizations.
}
\label{geo_3_2}
\end{figure}

In this section we present results of simulations of  a larger section of a mask.
Figure~\ref{geo_3_2}\,a) shows the layout, where red regions correspond to material
and green corresponds to blank regions.
The layout contains 65\,nm holes and posts (1X, 260\,nm in mask scale),
two Z-like structures with parameters corresponding to Layout~2.2 and further test structures.
Material and illumination and imaging parameters are listed in Table~\ref{table_material_stacks}
(material stack~1).

\begin{table}[h]
\begin{center}
\begin{tabular}{|l|r|r|r|r|l|l|}
\hline
\multicolumn{7}{|l|}{Layout 3 ($10\,\mu\mbox{m}~X~10\,\mu\mbox{m}$)}\\
\hline
 & p & t [s] & N & RAM [GB]& $I(0,0)$ & $I(1,1)$\\
\hline
 JCMsuite (FEM)
 &    3 & 1490 &1108800 &  & 0.6038803 & 0.0720966 \\
 &    3 &1637  &1206000 & 30 & 0.6043494 & 0.0720739 \\
 &    3 & 1967 &1335600 &  & 0.6044032 & 0.0721372 \\
 &    3 & 2588 & 1617600 & & 0.6043363 & 0.0719821 \\
 &    4 &1419  &1201200 &  & 0.6066360 & 0.0726171 \\
 &    4 & 1627 &1306500 & 33 & 0.6060792 & 0.0725927 \\
 &    4 & 1860 &1446900 &    & 0.6056927 & 0.0724837 \\
 &    4 & 2299 &1752400 &  47& 0.6050683 & 0.0723238 \\
 &    5 & 5037 &2217600 & 70 & 0.6057829 & 0.0721241 \\
 &    5 & 5524 &2412000 & 76 & 0.6054997 & 0.0721632 \\
 &    5 & 6182 &2671200 & 86 & 0.6053901 & 0.0721646 \\
&    5 & 8266 & 3235200 & 108  & 0.6054001 & 0.0722620 \\
\hline
Dr.LiTHO (WGM)
 &  & 20800 &  & 0.15  &0.6067974  & 0.0717071   \\
 \hline
\end{tabular}
\caption{Convergence of near field results for Layout~3. Computations with JCMsuite have been performed
using different finite element ansatz functions (parameter $p$) and different spatial discretizations of
the mask geometry (a finer mesh leads to a higher number of unknowns / expansion coefficients $N$).
RAM denotes the approximate memory consumption in GB.
JCMsuite computations have been performed using eight threads
of a multi-core workstation with extended RAM.
Dr.LiTHO computations have been performed using a single thread of a standard personal computer.
}
\label{table_large_results}
\end{center}
\end{table}

\begin{figure}[htb]
\centering
{\sffamily a)
\includegraphics[height=0.23\textwidth]{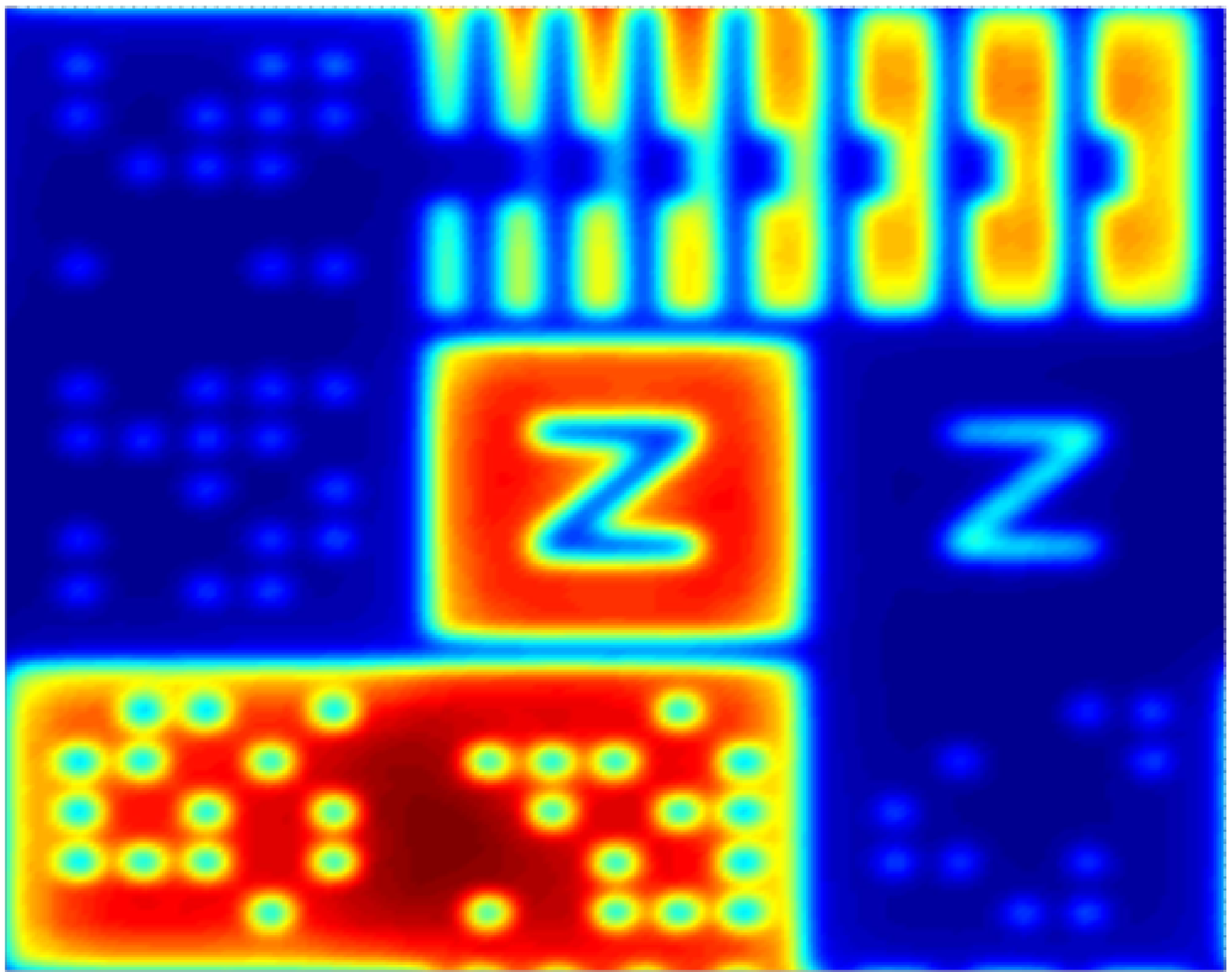}
b)
\includegraphics[height=0.23\textwidth]{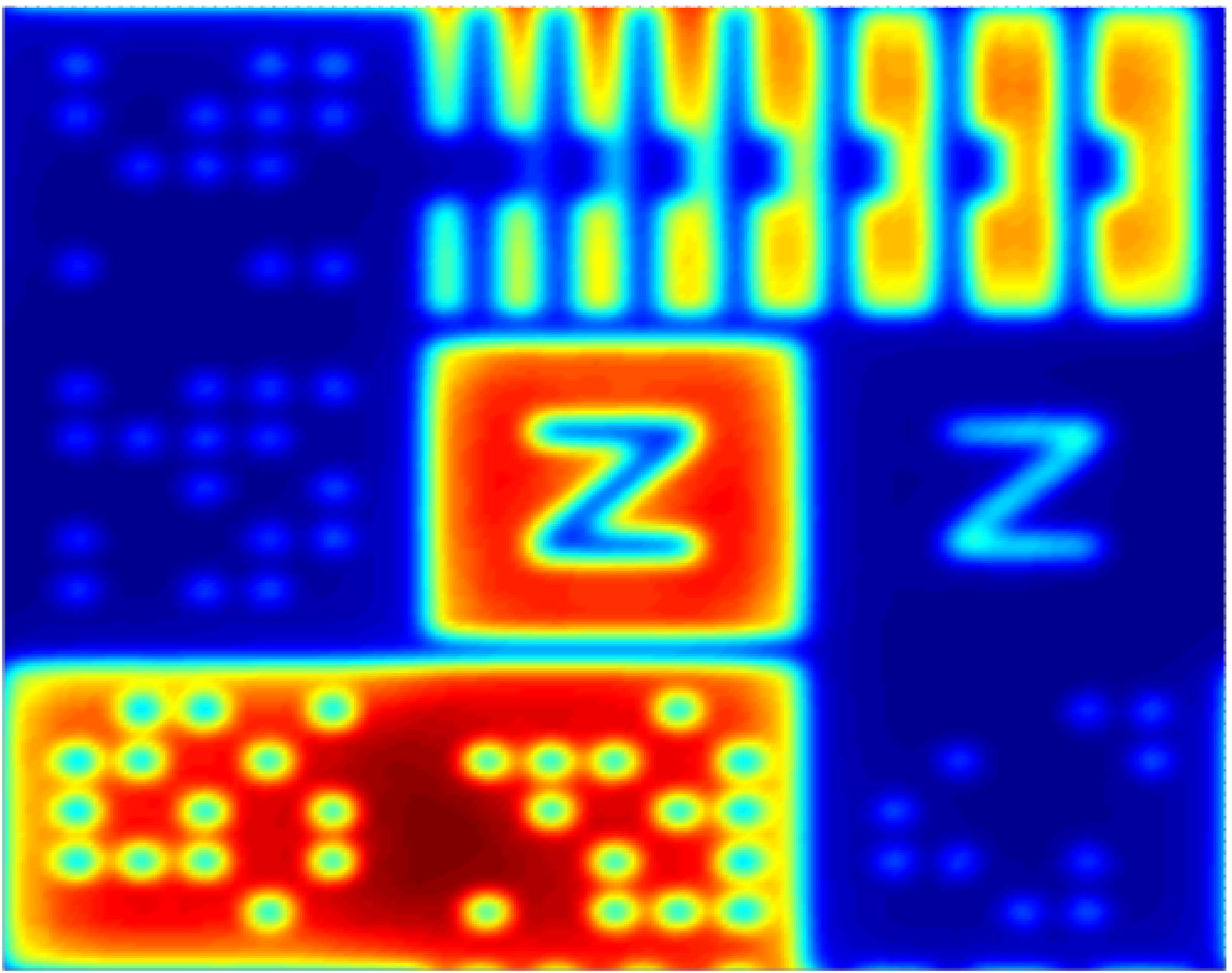}
c)
\includegraphics[height=0.23\textwidth]{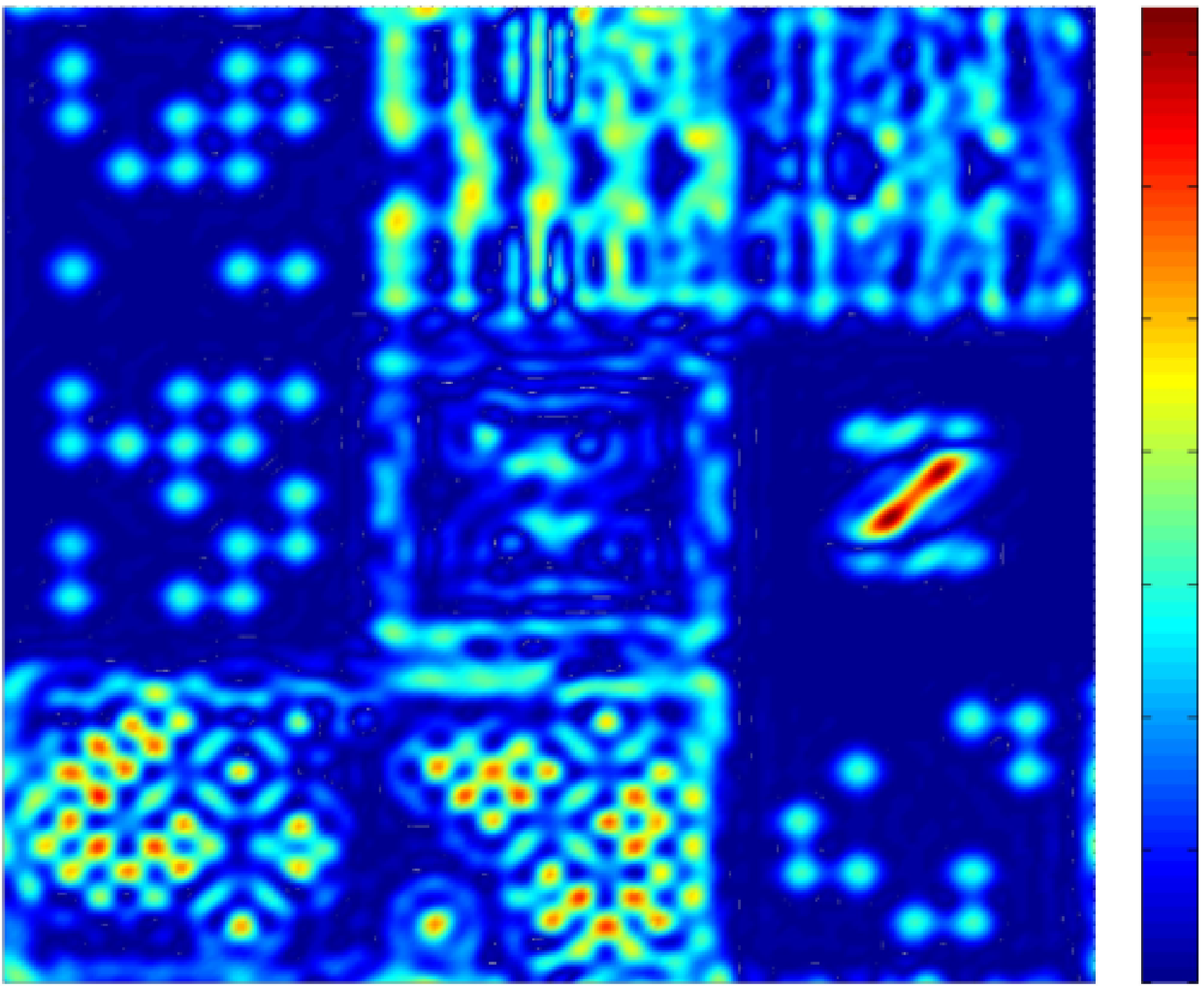}
}
\caption{
Aerial images from near fields computed with FEM (a) WGM (b).
c) Absolute value of the difference between a and b (pseudocolor range scales between zero and 1.5\% of the total intensity.)
}
\label{ai_3_2}
\end{figure}

\begin{figure}[htb]
\centering
{\sffamily
\includegraphics[width=0.4\textwidth]{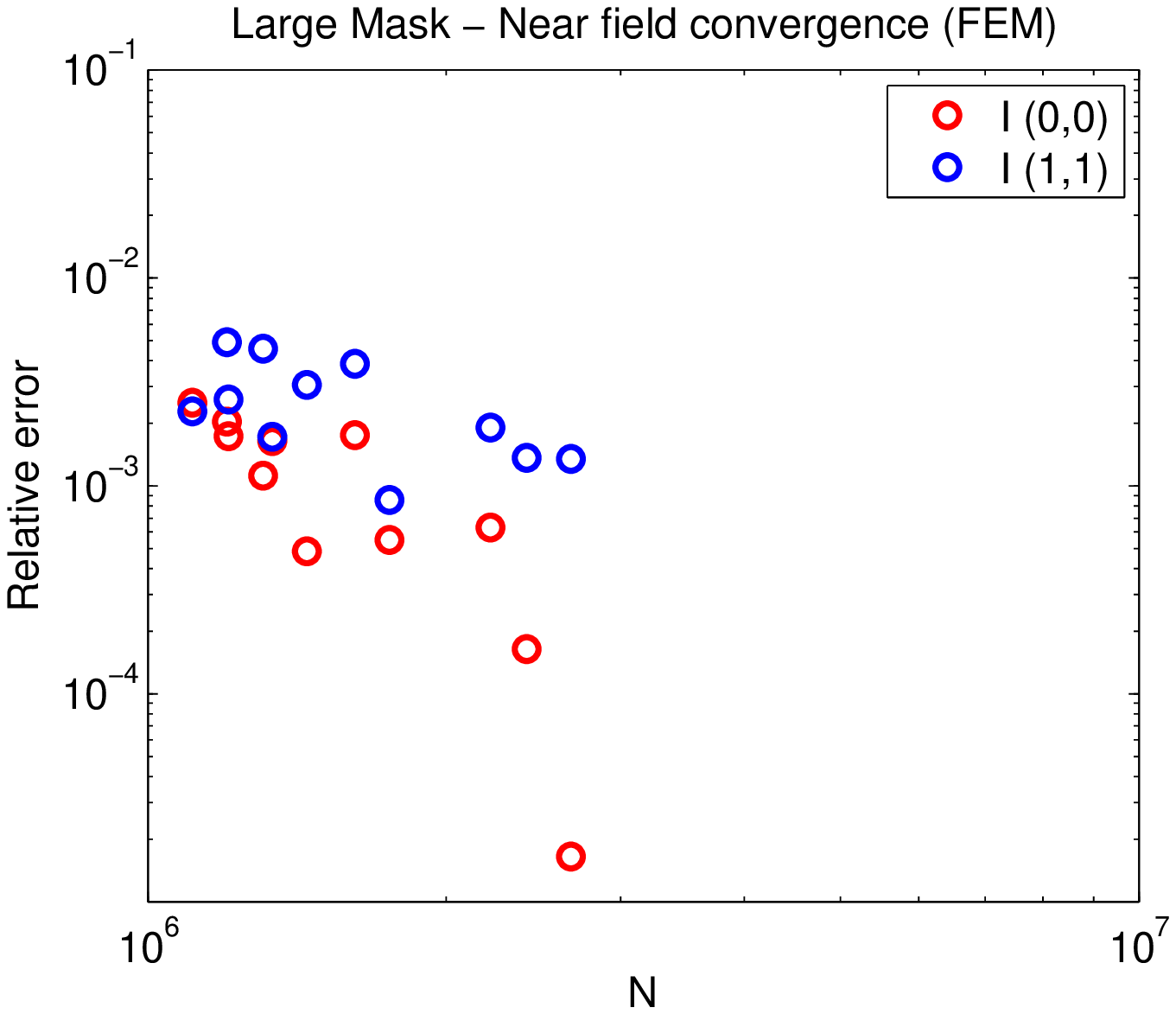}
}
\caption{
Convergence of the FEM near field results for Layout~3.1 ($10\,\mu m\,\times\,10\,\mu m$ mask) .
Dependence of the relative error of the magnitude of two distinct diffraction orders
on the number of unknowns in the finite-element problem, $N$.
(Compare Table~\ref{table_large_results}.)
 }
\label{conv_3}
\end{figure}

Table~\ref{table_large_results} shows the convergence of the near field results obtained with the
FEM solver for different numerical parameter settings. From the data for the diffraction order intensities
it can be seen that the relative error
of the significant diffraction efficiencies converges well to the range of about 0.1\%.
Convergence with finite element degree $p$ as well as convergence with mesh refinement, resp.~N, can
be observed.
Due to the large size of the problem we have not
investigated far field convergence in detail in this case.
However, the numerical settings in this case are similar to the settings in the previous chapters, i.e., we
expect the accuracy also of the far field results to be of the same quality as in Sections~\ref{contact_hole_chapter}
and~\ref{z_like_subsection} for the examples with the same material stacks.
This is also supported by the apparent good convergence of the near field results.
Further Table~\ref{table_large_results} shows computation time and memory consumption for WGM, for the
computation yielding the result displayed in Figure~\ref{ai_3_2}\,b). Here the domain decomposition
strategy of the solver has been applied (cf.~\ref{wgm_chapter}).
Please note that the computational effort in computation time is comparable (JCMsuite used eight threads in parallel
in this case, while Dr.LiTHO was used single-threaded) and that the computational effort in terms of
memory consumption is several orders of magnitude larger for FEM.

Figure~\ref{geo_3_2}\,b,c) shows near field distributions (at the upper boundary of the mask)
with source fields of different polarizations in a
pseudocolor representation. Interference fringes and field singularities at metal edges and corners can be
observed.
Figure~\ref{ai_3_2}\,a,b) shows corresponding aerial images, for near field computations with FEM and with WGM.
The images correspond very well to each other.
Plotting the differences of the fields (Figure~\ref{ai_3_2}\,c), one observes that difference of the aerial image
is not randomly distributed but localized especially around the Z features and at the 65\,nm contact holes and posts.
Please note the very good qualitative agreement of the
aerial image computed with WGM and with FEM.

\section{Conclusions}
\label{conclusions} We have developed an interface between the
lithography simulator Dr.LiTHO and the electromagnetic field solver
JCMsuite. We have demonstrated a typical lithography simulation flow
from the near field simulation to resist image formation,
post-exposure bake and development. Further, FEM-based JCMsuite and
a WGM-based solver from the package Dr.LiTHO have been benchmarked.
Both solvers have specific advantages and resulting from that
specific fields of applications. For standard lithography structures
WGM is more qualified. For more complex structures and for cases
where a very high accuracy is required FEM is more qualified.
Therefore the combination of Dr.LiTHO and JCMsuite forms a very
efficient lithography simulation environment for all fields of
applications.

\bibliography{/home/numerik/bzfburge/texte/biblios/phcbibli,/home/numerik/bzfburge/texte/biblios/group,/home/numerik/bzfburge/texte/biblios/lithography}
\bibliographystyle{spiebib}

\end{document}